\documentclass[aps,twocolumn,showpacs,pra,superscriptaddress,floatfix,longbibliography,nofootinbib]{revtex4-1}

\usepackage{amsmath,amsthm,amsfonts,amssymb,times,bbm,graphicx}
\usepackage{color,tabularx,epsfig,wasysym,dcolumn,bm,subfigure,psfrag}

\newcommand{\bra}[1]{\mbox{$\langle #1 |$}}
\newcommand{\ket}[1]{\mbox{$| #1 \rangle$}}

\newcommand{\R}{{\mathbb{R}}}

\begin{document}

\title{Pinning of Fermionic Occupation Numbers: General Concepts and One Dimension}

\author{Felix Tennie}
\affiliation{Clarendon Laboratory, University of Oxford, Parks Road, Oxford OX1 3PU, United Kingdom}
\author{Daniel Ebler}
\affiliation{Department of Computer Science, The University of Hong Kong, Pokfulam Road, Hong Kong}
\author{Vlatko Vedral}
\affiliation{Clarendon Laboratory, University of Oxford, Parks Road, Oxford OX1 3PU, United Kingdom}
\affiliation{Centre for Quantum Technologies, National University of Singapore, 3 Science Drive 2, Singapore 117543}
\author{Christian Schilling}
\email{christian.schilling@physics.ox.ac.uk}
\affiliation{Clarendon Laboratory, University of Oxford, Parks Road, Oxford OX1 3PU, United Kingdom}

\date{\today}

\begin{abstract}
Analytical evidence for the physical relevance of generalized Pauli constraints (GPCs) has recently been provided in [PRL 110, 040404]: Natural occupation numbers $\vec{\lambda}\equiv (\lambda_i)$ of the ground state of a model system in the regime of weak couplings $\kappa$ of three spinless fermions in one spatial dimension were found extremely close, in a distance $D_{min}\sim \kappa^8$ to the boundary of the allowed region. We provide a self-contained and complete study of this quasipinning phenomenon. In particular, we develop tools for its systematic exploration and quantification. We confirm that quasipinning in one dimension occurs also for larger particle numbers and extends to intermediate coupling strengths, but vanishes for very strong couplings. We further explore the non-triviality of our findings by comparing quasipinning by GPCs to potential quasipinning by the less restrictive Pauli exclusion principle constraints. This allows us to eventually confirm the significance of GPCs beyond Pauli's exclusion principle.
\end{abstract}

\pacs{03.67.-a, 05.30.Fk, 05.30.Jp}

\maketitle

\section{Introduction and elementary concepts}\label{sec:intro}
Since its formulation in 1925, Pauli's exclusion principle \cite{Pauli1925} has played a crucial role in the understanding of various phenomena, such as the atomic structure and related spectral observations, the stability of matter (see e.g.~Refs. \cite{Dyson1967,lieb1990}) and neutron stars. Only one year after its discovery, Heisenberg and Dirac recognized Pauli's exclusion principle to be a consequence of the more substantial fermionic exchange symmetry arising due to the indistinguishability of identical particles \cite{Dirac1926,Heis1926}. In terms of natural occupation numbers (NONs) $\lambda_i$, the eigenvalues of the 1-particle reduced density operator, Pauli's exclusion principle can be stated as
\begin{equation}\label{eq:PEP}
0 \leq \lambda_i \leq 1\,,\quad \forall i\,.
\end{equation}
Here, the NONs are normalized to the particle number N, $\lambda_1+\ldots+\lambda_d=N$ and we assume that the 1-particle Hilbert space $\mathcal{H}^{(d)}$ is finite, d-dimensional.
From a geometrical viewpoint, by ordering the $\lambda_i$ decreasingly and introducing the $\lambda$-vector $\vec{\lambda} \equiv (\lambda_i)_{i=1}^d$ and $\|\vec{x}\|_1\equiv \sum_{i=1}^d|x_i|$, Eq.~(\ref{eq:PEP}) restricts such vectors of NONs to the Pauli simplex $\Sigma$,
\begin{equation}\label{eq:Sigma}
\Sigma \equiv \{\vec{\lambda}\in \R^d\mid\|\vec{\lambda}\|_1=N\,,\,1\geq\lambda_1\geq\ldots\geq\lambda_d\geq 0 \}\,.
\end{equation}

In a number of works \cite{Borl1972,Kly2,Kly3,Altun,Rus2} the antisymmetry of the N-fermion wave function was found and proven only recently to impose a family of greater restrictions on $\vec{\lambda}$:
\begin{equation}\label{eq:gpc}
D_j(\vec{\lambda}) \equiv \kappa_j^{(0)}+\vec{\kappa}_j \cdot  \vec{\lambda}\,\geq \,0\,,\quad j=1,2,\ldots,r_{N,d},
\end{equation}
with $r_{N,d} < \infty$. Note that $(\kappa_j^{(0)},\vec{\kappa}_j)\in \mathbb{Z}^{d+1}$ as well as the number of constraints $r_{N,d}$ depend on the number of fermions N and the dimension d of the underlying 1-particle Hilbert space. It should be stressed that this recent breakthrough by Klyachko and Altunbulak \cite{Kly3,Kly2,Altun} was part of a more general effort in mathematical physics and quantum information theory \cite{Muel,Higushi,Brav,Kly4,Daft,MC,Liu,HallQMP,EisertGaussian,CSthesis,Vlach1,Alex,Vlach2} addressing the quantum marginal problem. This problem explores and describes the relations between reduced density operators (marginals) of subsystems arising from a common multipartite quantum state. One of the most prominant examples is the (2-body) N-representability problem which is about describing the set of 2-particle reduced density operators being compatible to N-fermion quantum states \cite{Col}.

\begin{figure}
\centering
\includegraphics[width=4.3cm]{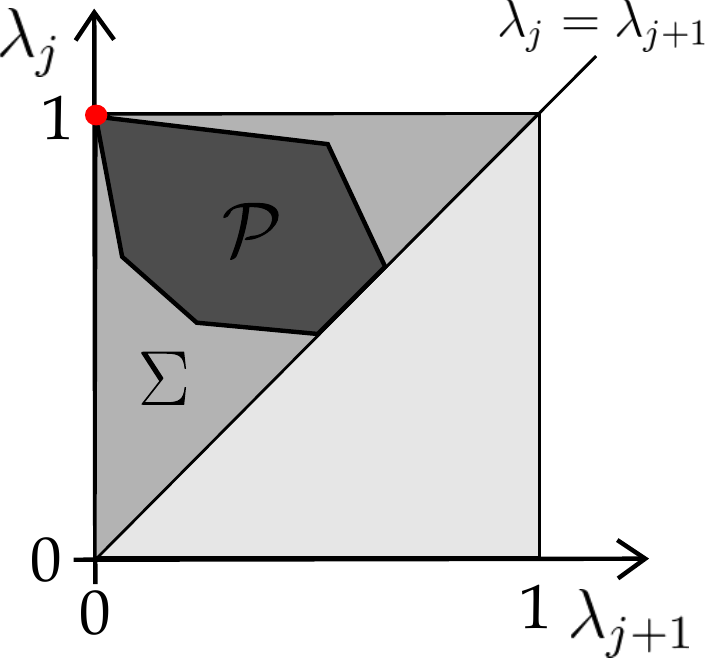}
\hspace{0.6cm}
\includegraphics[width=3.6cm]{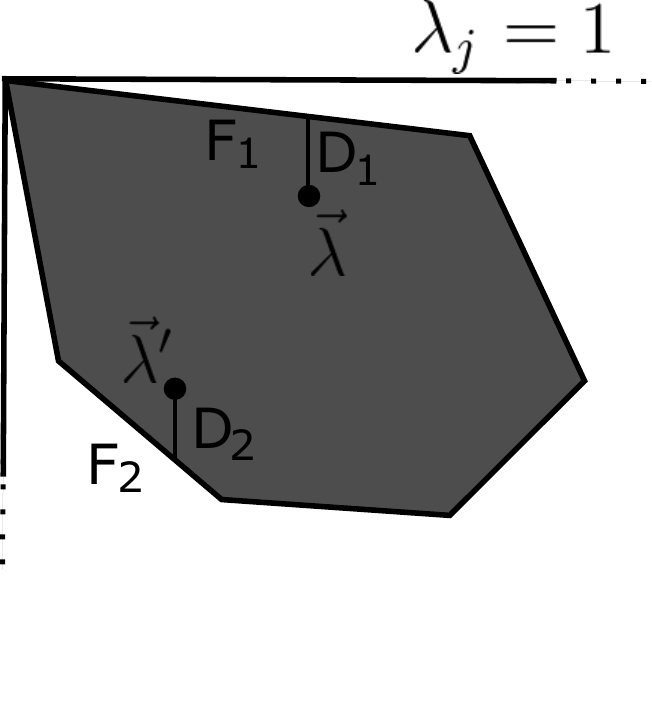}
\caption{Left: Schematic illustration of the (dark-gray) polytope $\mathcal{P}$ of possible vectors $\vec{\lambda}\equiv(\lambda_i)$ of decreasingly ordered NONs. $\mathcal{P}$ is a proper subset of the medium-gray `Pauli simplex' $\Sigma$ within the light-gray hypercube $[0,1]^d$. Right: Minimal $l^1$-distances $D_{min} = D_1$, $D_2$ of two different vectors $\vec{\lambda}$ and $\vec{\lambda}'$ to the polytope facets $F_1$ and $F_2$ are depicted (see Eq.~\eqref{eq:DefDmin} and text for more details).}
\label{fig:polytope}
\end{figure}

The so-called \emph{generalized Pauli constraints} (GPCs) \eqref{eq:gpc} determine a polytope-shaped subset $\mathcal{P}$ (see also Fig.~\ref{fig:polytope}),
\begin{equation}\label{eq:inclusions}
\mathcal{P} \,\subsetneq \,\Sigma \,\subset \,[0,1]^d\,.
\end{equation}
In other words, a $\lambda$-vector of NONs is compatible to a pure N-fermion quantum state $\ket{\Psi}\in \wedge^N[\mathcal{H}^{(d)}]$ if and only if $\vec{\lambda}$ lies in the polytope $\mathcal{P}$.
Here and in the following we typically suppress the dependence of $\mathcal{P}$ and $\Sigma$ on $N,d$.

Given the remarkable result on the GPCs, there is little doubt that these constraints will have some physical relevance as well.
For instance, from a general viewpoint, the GPCs may lead to new insights in reduced density matrix functional theory (RDMFT): Usually the minimization of a functional of the 1-particle reduced density operator to determine the ground state is erroneously considered to be only constrained by (\ref{eq:PEP}). Recently, it has been demonstrated for the first time that the GPCs can have a strong influence on the results of the minimization process for several functionals \cite{RDMFT}. In addition, the concept of master equations describing the dynamics of NONs may be modified by taking the GPCs into account. The GPCs might be also useful in tomography used for the reconstruction of the 1-particle reduced density matrix given some 1-particle information.

A more specific but potentially quite spectacular relevance of GPCs was postulated by Klyachko \cite{Kly1,Kly5} in the form of the \emph{pinning} effect: For some systems --- from the viewpoint of the 1-particle picture --- the ground state minimization process of the energy expectation value $\bra{\Psi_N}\hat{H}\ket{\Psi_N}$ for a Hamiltonian $\hat{H}$ might get stuck on the boundary of the polytope $\mathcal{P}$ since any further minimization would violate some GPC (\ref{eq:gpc}). Yet, in a first analytic investigation strong evidence was found for \emph{quasipinning} \cite{CS2013}. There, for the ground state of a few-fermion system the NONs were approximately saturating some GPC, $D_j(\vec{\lambda})\approx 0$, and therefore $\vec{\lambda}$ was found very close to, but not exactly on, the boundary of $\mathcal{P}$.

The occurrence of (quasi-)pinning gives rise to a number of important structural implications, such as a reduced complexity of the N-fermion wave function under expansion in terms of Slater determinants \cite{Kly1,CSthesis,CSQMath12} and a constrained dynamical evolution of the system \cite{CSQMath12,CS2015Hubbard}.

Over the past few years the study of GPCs, and the search for systems exhibiting (quasi)pinning in particular has therefore become a subject of growing interest among many branches of physics and quantum chemistry \cite{Kly1,BenavLiQuasi,Mazz14,CSthesis,CSQuasipinning,BenavQuasi2,RDMFT,Alex,CS2015Hubbard}.
Yet most of those works resorted to numerical methods, and in addition employed quite strong approximations: The 1-particle Hilbert space was truncated from infinite dimensions to at most six up to eight. As an unfortunate consequence, the NONs of the approximated ground states turn out to differ quite a lot from those of the \emph{correct} ground state and no conclusive statement on the occurrence of (quasi)pinning for the correct ground state was possible \cite{ETHprivate}. This also renews the caveat already expressed in \cite{CS2013}: \emph{``it is likely extremely challenging to use numerical methods to distinguish between genuinely pinned and mere quasipinned states. This underscores the need for analytical analyses,\ldots''}.
Moreover, little if anything has been understood so far about the origin of (quasi-)pinning.

The present paper thus aims to shed light onto these open questions and aspects. A self-contained description of how to investigate quasipinning systematically is provided. In a comprehensive analysis, the scope of quasipinning is explored with respect to different particle numbers and varying coupling strengths. The paper is organized as follows. In Section \ref{sec:concepts} the concept of a systematic \mbox{(quasi-)pinning} analysis is explained, accompanied with some insights on the polytope structure/GPCs. Section \ref{sec:Q} stresses that quasipinning in some cases is trivial (e.g.~as a consequence of weak correlations) and discusses a measure for distinguishing non-trivial from such trivial quasipinning. The physical model and its key characteristics are defined and outlined in Section \ref{sec:model}.
In the main section, Section \ref{sec:1d}, the occurrence of quasipinning is analysed for this Harmonium-model in one spatial dimension for arbitrary particle numbers, from weak, intermediate even up to very strong couplings.

\section{Quasipinning measure and concept of truncation}\label{sec:concepts}
In this section we elaborate on possible measures for quasipinning and provide some geometric
insights on the polytope defined by the generalized Pauli constraints (GPCs).

\subsection{Quasipinning measures}\label{sec:QPmeasures}
First of all, since each setting of N particles and dimension d of the 1-particle Hilbert space gives rise to several GPCs, $D_j$, $j=1,2,\ldots,r_{N,d}$, the information of (quasi)pinning might be further specified by stating the corresponding GPC showing (quasi)pinning. This is expressed geometrically: If a given vector $\vec{\lambda}$ of NON saturates a GPC $D_j(\cdot)\geq 0$ we say that the corresponding vector of NON \emph{is pinned to the corresponding facet $F_{D_j}$} of the polytope $\mathcal{P}$ which is defined by
\begin{equation}
F_{D_j}\equiv \{\vec{\lambda}\in\mathcal{P}\,\mid\,D_j(\vec{\lambda})=0\}\,.
\end{equation}
To quantify the strength of quasipinning one is tempted to choose just the value $D(\vec{\lambda})$. Yet, this involves a subtlety. Since any GPC $D(\cdot)\geq 0$ is equivalent to $\alpha D(\cdot)\geq 0$ for any $\alpha>0$ there is an ambiguity which can be fixed by
expressing each GPC (\ref{eq:gpc}) in its canonical form. This form is given by choosing the minimal possible integer coefficients $\kappa^{(0)},\ldots,\kappa^{(d)}$ \footnote{That one can choose all coefficients $\kappa^{(i)}$ as integers is a non-trivial mathematical fact \cite{Kly2}.}. In that way, we have defined for each GPC a corresponding natural quasipinning measure given by the value $D_j(\vec{\lambda})$.

Alternatively, the geometric structure in the form of a polytope suggests to choose as quasipinning measure the $l^p$-distances of $\vec{\lambda}$ to the corresponding facets $F_{D_j}$ for some $p$. Since the 1-particle reduced density operator is normalized with respect to the trace,
\begin{equation}\label{eq:1RDOnorm}
\mbox{tr}[\rho]=\sum_{i=1}^d \lambda_i = \|\vec{\lambda}\|_1 \overset{!}{=} N\,,
\end{equation}
the $l^1$-norm seems to be the most obvious one. It turns out to be closely related to the natural measure given by $D(\vec{\lambda})$ (see Supplemental Material of Ref.~\cite{CSQ}),
\begin{equation}\label{eq:l1}
\mbox{dist}_1(\vec{\lambda},E_D)= 2 \, D(\vec{\lambda})\,.
\end{equation}
Here, $E_D$ denotes the hyperplane obtained by extending $F_D$ to `points' $\vec{\lambda}$ outside of the polytope $\mathcal{P}$ (but still normalized to N).
Consequently, as long as the minimal $l^1$-distance of $\vec{\lambda}$ to $E_D$ is attained within the polytope, $\mbox{dist}_1(\vec{\lambda},F_D)$ coincides with $D(\vec{\lambda})$ (up to a factor `2'). For any other $p$, relations for $\mbox{dist}_p(\vec{\lambda},E_D)$ of the same form as (\ref{eq:l1}) can be found. The prefactor of 2, however, is replaced by a specific function of the coefficients $\kappa^{(i)}$ (depending on $p$). The independence of the factor `2' on $\{\kappa^{(i)}\}$ in Eq.~(\ref{eq:l1}) for $p=1$ also makes the use of the $l^1$-distance preferable.

Further insights on the choice of the most significant quasipinning measure can only be obtained by understanding the potential physical relevance of quasipinning. To briefly comment on that, recall that pinning of $\vec{\lambda}$, as an effect in the 1-particle picture, allows one to reconstruct the structure of the corresponding N-fermion quantum state $\ket{\Psi_N}\in \wedge^N[\mathcal{H}^{(d)}]$. In addition, $\ket{\Psi_N}$ is significantly simplified since it is given by a linear combination of only a few, specific Slater determinants (for details we refer the reader to Refs.~\cite{Kly1,CSthesis,CSQMath12,CSQuasipinning}). In case of quasipinning the same holds for $\ket{\Psi_N}$ up to a small error. First results provided in Ref.~\cite{CSQuasipinning} show that this error is bounded linearly in $D(\vec{\lambda})$ from above (and also from below). These results suggest $D(\vec{\lambda})$ to be the most significant quasipinning measure.

Since there are quite a few GPCs for settings with $N\geq 4$ and $d\geq 8$
we also define the overall quasipinning measure by
\begin{equation}\label{eq:DefDmin}
D_{min} \equiv \min_{j} \left[D_j(\vec{\lambda})\right].
\end{equation}
Up to a prefactor, $D_{min}$ thus resembles the distance measure $\mbox{dist}_1(\cdot,\cdot)$ with respect to the $l^1$-norm between the $\lambda$-vector and the boundary $\partial\mathcal{P}$ \footnote{By $\partial \mathcal{P}$ we only refer to that part of the boundary of the polytope $\mathcal{P}$ which corresponds to saturation of some GPC. The remaining part of the polytope boundary described by saturation of an ordering constraints $\lambda_i-\lambda_{i+1}\geq 0$ is not relevant here. In particular, notice that the saturation of an ordering constraint does not in general lead to any directly accessible simplification for the N-fermion quantum state $\ket{\Psi_N}$.}.

\subsection{Concept of truncation}\label{sec:truncation}
For quasipinning analyses in practice one faces a major problem. On the one hand, the complete family of GPCs is known so far only for the settings $(N,d)$ up to $d=10$ \cite{Altun} and in addition also for the settings $(3,11),(8,11)$ \cite{MApriv}. On the other hand, most few-fermion models are based on an infinite-dimensional 1-particle Hilbert space given by $\mathcal{H}\equiv L^2(\mathcal{C})$ since the 1-particle configuration space $\mathcal{C}$ is typically continuous (e.g.~$\mathcal{C}=\mathbb{R}^3$).

A common (see Refs.~\cite{Kly1,BenavLiQuasi,Mazz14,BenavQuasi2,RDMFT}), but less reasonable way to circumvent that problem is to truncate $\mathcal{H}$ from the very beginning to just $d=10$ or even less dimensions (i.e.~to at most 5 orbitals in the case of electrons) and restrict the Hamiltonian to the corresponding $\binom{d}{N}$-dimensional subspace $\wedge^N[\mathcal{H}^{(d)}]$. Unfortunately, for most physical models this drastic approximation does not allow one to conclusively explore the occurrence of quasipinning for the exact ground state. Besides the objection that the system may be in general too correlated in order to justify such a truncation also a less optimal (erroneous) choice for the truncated $\mathcal{H}^{(d)}$ can lead to wrong results on quasipinning. This is even the case for weakly correlated systems, such as atoms.

A systematic way to avoid the error related to the choice of $\mathcal{H}^{(d)}$ is to implement such a truncation to a small d \emph{after} having obtained a sufficiently accurate approximation for the exact ground state. We briefly explain how this \emph{concept of truncation} works and discuss the underlying mathematical structure.
The main idea is that the pinning analysis for $\vec{\lambda}=(\lambda_i)_{i=1}^{d'}$ belonging to the setting $(N',d')$ (e.g., with $d'$ infinite) can be simplified by skipping various NON sufficiently close to 1 and 0. Then, possible quasipinning of the truncated vector containing the remaining NONs can be explored in the setting of smaller N and smaller d. The corresponding result on possible quasipinning in the truncated setting translates to quasipinning of the same strength in the larger setting up to a small error.
This is based on the following polytope relation (we reintroduce indices `$(N,d)$')
\begin{equation}\label{eq:Prelation}
\mathcal{P}_{N',d'}|_{
\tiny \begin{array}{l}
\lambda_1=\ldots =\lambda_r=1\\
\lambda_{d'+1-s}=\ldots =\lambda_{d'}=0
\end{array}
} = \,\mathcal{P}_{N,d}\,,
\end{equation}
where $0\leq r \leq N'$, $0\leq s \leq d'-N'$, $N\equiv N'-r$ and $d \equiv d'-r-s$.
In words, relation (\ref{eq:Prelation}) states that restricting the polytope $\mathcal{P}_{N',d'}$ to the hyperplane defined by $\lambda_1=\ldots =\lambda_r=1$,
$\lambda_{d'+1-s}=\ldots =\lambda_{d'}=0$ leads to the corresponding polytope for the setting
$(N'-r,d'-r-s)$ (which is embedded in the larger space $\mathbb{R}^{d'}$).

On the level of GPCs this implies that for every GPC $D_j$ of the smaller setting $(N,d)$ there exists \emph{at least} one so-called extended constraint $D'_j$ in the larger setting $(N',d')$, i.e.
\begin{equation}\label{eq:Drelation1}
D_j(\vec{\lambda}) = D'_j(\underbrace{1,\ldots,1}_r,\vec{\lambda},\underbrace{0,\ldots,0}_s)\,,\quad\forall \vec{\lambda}\in \mathcal{P}_{N,d}\,.
\end{equation}
For such associated constraints, the linearity of GPCs (\ref{eq:gpc}) implies
\begin{equation}\label{eq:Drelation2}
D'_j(\vec{\lambda}') = D_j(\vec{\lambda})+\mathcal{O}(1-\lambda'_r) + \mathcal{O}(\lambda'_{d'+1-s})\,,
\end{equation}
where $\vec{\lambda}\equiv (\lambda'_j)_{j=r+1}^{d'-s}$.

All the remaining constraints of the larger setting lead to restrictions in the smaller setting
which are linearly dependent on the GPCs of that smaller setting.

\begin{figure}[!h]
\includegraphics[width=8cm]{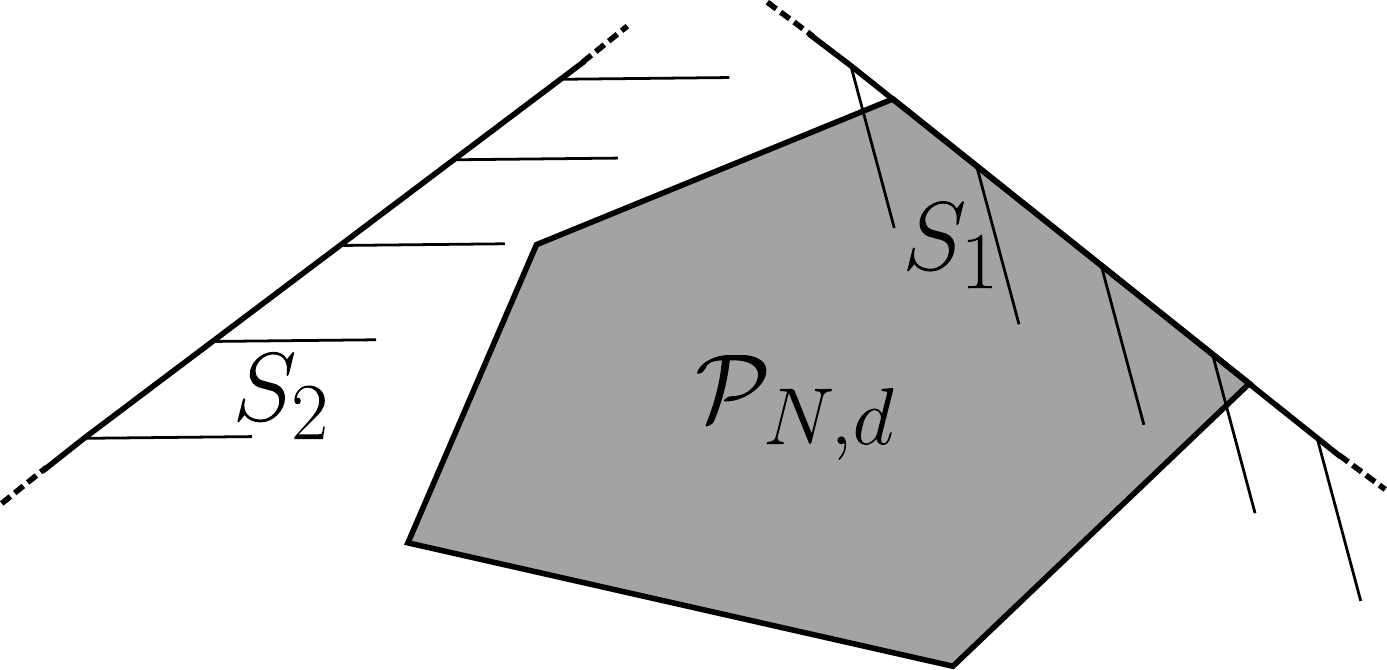}
\centering
\caption{Polytope $\mathcal{P}_{N,d}$ and illustration of half-spaces defined by generalized Pauli constraints of larger setting $(N',d')$ restricted to smaller setting $(N,d)$.}
\label{fig:facetproj}
\end{figure}

These two situations are also illustrated in Figure \ref{fig:facetproj},
where two half-spaces, $S_1$ and $S_2$, are shown. Each of them is defined by all vectors $\vec{x}\in \mathbb{R}^d$ fulfilling a specific GPC $D'$ of the larger setting after restricting it to the smaller setting $(N,d)$,
\begin{equation}
S = \{\vec{x}\in \mathbb{R}^d\,|\,D'(\underbrace{1,\ldots,1}_r,\vec{x},\underbrace{0,\ldots,0}_s)\geq 0\}\,,
\end{equation}
with $r\equiv N'-N$ and $s\equiv (d'-N')-(d-N)$.
For instance, the half-space $S_1$ in Figure \ref{fig:facetproj} corresponds to a proper GPC of the smaller setting. In contrast, the half-space $S_2$ does not describe a proper GPC in the smaller setting and depends linearly on some proper GPCs \footnote{Among the latter class of constraints $D'$ of the larger settings there are also a few taking even a simpler form after restriction to the setting $(N,d)$. They reduce to simple tautologies, \mbox{$D'(\underbrace{1,\ldots,1}_r,\vec{x},\underbrace{0,\ldots,0}_s)=\mbox{const}\geq 0$.}}.

Based on these insights on the polytope structure (\ref{eq:Prelation}), which led to Eq.~(\ref{eq:Drelation2}), the concept of truncation emerges: NONs sufficiently close to 1 or 0 can be neglected and the error of the truncated pinning analysis is given by Eq.~(\ref{eq:Drelation2}). In Section \ref{sec:1d} we will successfully make use of that concept to conclusively explore quasipinning for a system based on an \emph{infinite-dimensional} 1-particle Hilbert space.

The concept of truncation and relation (\ref{eq:Prelation}) in particular, also emphasizes the choice of $\{D_j(\vec{\lambda})\}$ and Eq.~(\ref{eq:DefDmin}) as the most natural quasipinning measures: It is a desirable property of such measures to lead to the same results on quasipinning
for vectors $\vec{\lambda} \in \mathcal{P}_{N,d}$ and $\vec{\lambda}' \in \mathcal{P}_{N',d'}$
differing only by 0's and 1's. This so-called \emph{truncation-consistency} is given for the natural measures $\{D_j(\vec{\lambda})\}$ and Eq.~(\ref{eq:DefDmin}) but not for distance measures given by the $l^p$-norm for $p >1$. The latter statement follows from the fact that for $p>1$ the prefactor on the right-handed side in Eq.~(\ref{eq:l1}) is not independent of the $\kappa^{(i)}$'s. By extending a given $\vec{\lambda}$ by 0's the values $D_i(\vec{\lambda})$ will not change but the prefactor will do (since the $\kappa^{(j)}$ for the extra dimensions will enter and can change it). In that case the $l^p$-distance will change although $\vec{\lambda}$ was extended only by `irrelevant' 0's.

\section{Generalized Pauli constraints beyond Pauli's exclusion principle constraints}\label{sec:Q}
The values $\{D_j(\vec{\lambda})\}$ (recall their canonical form as described in the previous section) and definition (\ref{eq:DefDmin}), respectively, provide natural measures for quasipinning.

Whenever quasipinning is very strong, i.e.~some $D_j(\vec{\lambda})$ are found to be very small, we can expect that the fermionic exchange symmetry becomes significant for the system from the 1-particle picture's viewpoint. Yet, such potential relevance of GPCs for many physical systems is not surprising at all. Due to the inclusion relation (\ref{eq:inclusions}), illustrated in Figure \ref{fig:polytope}, the relevance of GPCs already follows from the well-known relevance of the \emph{less restrictive} Pauli exclusion principle constraints: Whenever $\vec{\lambda} \in \mathcal{P}$ lies close to the boundary of the Pauli simplex $\Sigma$ it lies close to the boundary of $\mathcal{P}$ as well. For instance, for weakly correlated systems the vector of NONs lies close to the Hartree-Fock point given by $(1,\ldots,1,0,\ldots)$ (shown as `red dot' in Figure \ref{fig:polytope}). Therefore, it lies also close to the polytope boundary. Even strongly correlated fermionic quantum systems have typically some fermions (electrons) which are strongly bound, e.g., in atomic 1s shells. This quasipinning of the first Pauli exclusion principle constraints $1-\lambda_i\geq 0$, $i=1,2,\ldots$ geometrically implies quasipinning by GPC.

Consequently, the question is whether GPCs have any additional relevance, i.e.~\emph{beyond} Pauli's exclusion principle. Do given NONs $\vec{\lambda}$ lie significantly closer to any polytope facet than one could expect from a possibly small distance of $\vec{\lambda}$ to the boundary of the Pauli simplex (\ref{eq:Sigma})?

Recently, a measure for the degree by which quasipinning by GPCs exceeds that by the Pauli exclusion principle constraints has been introduced \cite{CSQ}, the so-called \emph{Q-parameters}. For each GPC $D_j$, $Q_j$ evaluates the ratio of the distances of  $\vec{\lambda}$ to the facet $F_j$ and the corresponding part of the boundary of the Pauli simplex. For the technical details and the properties of those Q-parameters we refer the reader to Ref.~\cite{CSQ}. A $\lambda$-vector with associated $Q_j(\vec{\lambda})$ lies $10^{Q_j}$ times closer to the facet $F_j$ of the polytope than could be expected from its possibly small distance to the boundary of the Pauli simplex $\Sigma$.

Similar to the definition (\ref{eq:DefDmin}) of an overall quasipinning measure we define $Q(\vec{\lambda})$ as the maximum of various $Q_j(\vec{\lambda})$. These concepts are visualised in Figure \ref{fig:polytope} where on the right minimal $l^1$-distances $D_{min} = D_1$, $D_2$ of two different vectors $\vec{\lambda}$ and $\vec{\lambda}'$ to the polytope facets $F_1$ and $F_2$ are shown. In contrast to $\vec{\lambda}$, the Pauli exclusion principle constraint is not approximately saturated by $\vec{\lambda}'$ which leads to $Q(\vec{\lambda}') > Q(\vec{\lambda})$.

We like to conclude this section by commenting on the term `trivial quasipinning'. First of all, quantifying quasipinning on an absolute scale by $\{D_j(\vec{\lambda})\}$ and $D_{min}(\vec{\lambda})$, respectively, has its own justification. In that way, one can explore
the absolute influence of the exchange symmetry in the 1-particle picture. Yet, to emphasize the potential indispensable necessity of the concept of GPCs one needs to
explore in addition whether quasipinning by GPCs exceeds that by Pauli's exclusion principle or not. In the following we will always carefully distinguish between such non-trivial and trivial
quasipinning, as quantified by the Q-parameter.

\section{Model}\label{sec:model}
In this section we introduce an analytically solvable few-fermion model and explain how to calculate the NONs of its ground state. This particularly includes the technical details of Ref.~\cite{CS2013}.

In order to investigate the influence of the fermionic exchange symmetry from the 1-particle picture's viewpoint it is particularly instructive to choose a model system that exhibits a considerable conflict between the energy minimization and the antisymmetry.
The system of N harmonically interacting fermions in a harmonic external trapping potential provides these features since the particle-particle interaction diminishes with decreasing particle distance. Accordingly, it is not the pair interaction which prevents the particles from sitting on top of each other but solely the fermionic exchange symmetry.
The corresponding Hamiltonian reads as
\begin{align}\label{eq:HamiltonianHarmonium01}
H_N = \sum_{i=1}^{N} \left(\frac{p_i^{\,2}}{2 m} +\frac{m}{2} \omega^2 {x_i}^2\right) + \frac{K}{2} \sum_{1\leq i<j\leq N} (x_i-x_j)^2,
\end{align}
where  $p_i$ and $x_i$ represents the momentum and position operators of the i-th particle. In the present paper this model shall be
referred to as Harmonium\footnote{In the literature, the term Harmonium has been used for two different systems: a) N harmonically interacting particles in a harmonic external trapping potential (see e.g.~Refs.~\cite{HarmName1,Benavides:fermVsBos}) b) N particles interacting via Coulomb forces in a harmonic external trapping potential (see e.g.~\cite{Cioslowski2006}); we shall follow convention a).}.

The choice of this particular model offers additional advantages: First, being analytically solvable, it allows us to gain structural insights and has therefore been chosen for thorough investigation in this paper.
Second, the model, which is also known as a Moshinsky-type atom, has seen interest and successful application in various branches of physics and physical chemistry, such as atomic physics, quantum dots, quantum information theory and entanglement theory \cite{Bouvrie2012,Johnson1991,Yauez2010,Laguna2011,Laguna2012,Dahl2009,Nagydual1,Nagy2011a,Nagy2012,Nagy2013,CS2013NO,Nagydual2,Benavides:fermVsBos,koscik2015neumann}. This may trigger crosslinks and will permit our results to be used in the context of other disciplines as well.

We explore the characteristic behaviour of (quasi-)pinning in various coupling regimes
by studying cases of different particle numbers N. Such a comprehensive investigation will also allow us to shed light onto the origin of quasipinning.

\subsection{Fermionic ground state}\label{sec:gs}
\emph{A priori}, Hamiltonian \eqref{eq:HamiltonianHarmonium01} acts as an operator on the N-particle Hilbert space $\mathcal{H}_N = \otimes_{i=1}^{N}\mathcal{H}$, where the 1-particle Hilbert space has been denoted by $\mathcal{H}$. Here, $\mathcal{H}$ is given by $\mathcal{H}\equiv L^2(\R)$.

$H_N$ is invariant under any permutation of particles. This particularly allows us to consider (\ref{eq:HamiltonianHarmonium01}) as a Hamiltonian for identical fermions (or bosons). In case of fermions we need to restrict (\ref{eq:HamiltonianHarmonium01}) to the subspace of fermionic quantum states, given by all states being antisymmetric,
\begin{equation}\label{eq:subspaceF}
\mathcal{H}_N^{(f)} \equiv \wedge^N[\mathcal{H}] \lneq  \mathcal{H}_N \equiv \mathcal{H}^{\otimes^N} \,.
\end{equation}

In order to derive the set of \emph{fermionic} eigenstates of \eqref{eq:HamiltonianHarmonium01} one may therefore initially derive the set of all N-particle eigenstates on $\mathcal{H}_N$ followed by a projection onto the fermionic subspace $\mathcal{H}_N^{(f)}$.
In Ref.~\cite{HarmChin} the fermionic spectrum and corresponding eigenstates were determined. For the ground state one finds
\begin{eqnarray}\label{eq:gsf}
\Psi^{(f)}(\vec{x}) &=& \mathcal{N}   \cdot  \Big[\!\prod_{1\leq i<j\leq N}\!(x_i-x_j)\Big] \cdot \exp{\left[-\frac{1}{2 \tilde{l}^2} \vec{x}^2\right]}\\
&&\cdot   \exp{\left[\frac{1}{2N}\left(\frac{1}{\tilde{l}^2} -\frac{1}{l^2}\right) (x_1+\ldots+x_N)^2\right]}\,,\nonumber
\end{eqnarray}
where $\vec{x}\equiv (x_i)_{i=1}^N$ and $\mathcal{N}$ is a normalisation constant. Here,
$l \equiv \sqrt{\frac{\hbar}{m\omega}}$ denotes the natural length scale for the center of mass motion and
$\tilde{l} \equiv \sqrt{\frac{\hbar}{m \omega \sqrt{1+ N K/(m\omega^2)}}}$ that for the relative motion. It is also worth noticing that the fermionic ground state (\ref{eq:gsf}) differs from the bosonic ground state only by the additional polynomial prefactor, the Vandermonde determinant.

Finally, it should be emphasized that the specific structure of the Hamiltonian (\ref{eq:HamiltonianHarmonium01}) implies that the NONs of the fermionic ground state (and of any other eigenstate) do not depend on $m \omega^2$ and $K$ separately but just on their ratio. This suggests the following
definition of a dimensionless coupling strength
\begin{equation}\label{eq:kappa}
  \kappa \equiv \frac{N K}{m \omega^2} = \left(\frac{l}{\tilde{l}}\right)^4-1\,.
\end{equation}

\subsection{1-particle reduced density operator}\label{sec:1rdm}
To determine the fermionic 1-particle reduced density operator $\rho(x,y)$ (in spatial representation) we need to integrate out $N-1$ fermions. For fixed particle number, as an exercise in Gaussian integration, one finds \cite{CS2013NO}
\begin{equation}\label{eq:1rdof}
\rho^{(f)}(x,y) = F(x,y)\,e^{-\alpha (x^2+y^2) +\beta x y}\,.
\end{equation}
where the symmetric polynomial $F$ and the parameters $\alpha,\beta$ depend on N and the coupling strengths
$m\omega^2, K$ and the length scales $l, \tilde{l}$, respectively. $\rho^{(f)}(x,y)$ coincides with the 1-particle reduced density operator (see Ref.~\cite{CS2013NO})
\begin{eqnarray}\label{eq:1rdob1}
\rho^{(b)}(x,y) &=& \sqrt{\frac{2\alpha-\beta}{\pi }}\, e^{-\alpha (x^2+y^2) +\beta x y}
\end{eqnarray}
of the bosonic ground state up to the polynomial prefactor $F(x,y)$, originating from the Vandermonde determinant in Eq.~(\ref{eq:gsf}).

Notice that $\rho^{(b)}(x,y)$ is the Euclidean Feynman propagator for the harmonic oscillator and can therefore be diagonalized,
\begin{equation}\label{eq:1rdob2}
\rho^{(b)}(x,y) = c(\kappa)\,\sum_{k=0}^\infty\,\frac{q(\kappa)^k}{k!}\,\varphi_k^{(L)}(x)\,\varphi_k^{(L)}(y)\,.
\end{equation}
Here, $c(\kappa)\equiv N\,\big(1-q(\kappa)\big)$ and the bosonic natural orbitals, i.e.~the eigenstates of $\rho^{(b)}(x,y)$, are given by the Hermite functions $\{\varphi_k^{(L)}(x)\}_{k=0}^{\infty}$ with natural length scale \mbox{$L\equiv \sqrt{l \,\tilde{l}}\,\left[\frac{(N-1){\tilde{l}}^{2} +l^2}{\tilde{l}^2 +(N-1)l^2}\right]^{\frac{1}{4}}$} \cite{CS2013NO}. The decay factor $q(\kappa)$ can easily be calculated by using the results in Ref.~\cite{CS2013NO} and one obtains
\begin{equation}\label{eq:qkappa}
q(\kappa) = 1 - \frac{2N}{N+\sqrt{N^2-(N-1)\left[2-(1+\kappa)^2-1/(1+\kappa)^2\right]}}\,.
\end{equation}

In contrast to $\rho^{(b)}(x,y)$, $\rho^{(f)}(x,y)$ cannot be diagonalized analytically.
Yet we can diagonalize $\rho^{(f)}(x,y)$ for given N either by numerical means for fixed couplings or by a perturbative approach for the regime of weak couplings. As a first step for this we need to map $\rho^{(f)}(x,y)$ as a density kernel to a matrix. The similarity between $\rho^{(f)}(x,y)$ and $\rho^{(b)}(x,y)$ strongly suggest to choose the bosonic natural orbitals $\{\varphi_k^{(L)}(x)\}$ as a reference basis. The corresponding  matrix
\begin{equation}\label{eq:1rdm}
(\rho^{(f)}_{kn})\equiv \big(\langle\varphi_k^{(L)}|\rho^{(f)}|\varphi_n^{(L)}\rangle\big)_{k,n\geq0}
\end{equation}
has a simplified form based on the following three properties:
\begin{enumerate}
\item Since the polynomial prefactor $F(x,y)$ is of finite degree $2(N-1)$ and since $x,y$ can be replaced by ladder operators with respect to the harmonic oscillator states with length scale $L$, $(\rho^{(f)}_{k n})$ is a band matrix, i.e.
    \begin{equation}\label{eq:1rdmband}
    \rho^{(f)}_{k n} = 0 \quad,\,\mbox{whenever}\,|k-n|>2(N-1)\,.
    \end{equation}
\item Since Hamiltonian (\ref{eq:HamiltonianHarmonium01}) has a 1-particle symmetry,
$[H_N,U(P)^{\otimes^N}]=0$, given by the simultaneous reflections $P:x_i\rightarrow -x_i$, the ground state and its 1-particle reduced density operator inherit this symmetry \cite{Dav}. Since $[\rho^{(f)},U(P)]=0$ we have
\begin{equation}\label{eq:1rdoP}
    \rho^{(f)} = \rho_{even}^{(f)} \oplus \rho_{odd}^{(f)}\,
\end{equation}
where `$even$' and `$odd$' stand for the corresponding parity of $U(P)$ (recall $U(P)^2 = \mathbf{1}$).
Moreover, since each reference basis state $\varphi_k^{(L)}$ also respects that symmetry, $U(P)\varphi_k^{(L)}= (-1)^k \varphi_k^{(L)}$, we find
in particular $(\rho^{(f)}_{k n}) =0$ whenever $k+n$ odd.
\item The dominating exponential factor in (\ref{eq:1rdof}), also leading to the decaying factor $q(\kappa)<1$ in Eq.~(\ref{eq:1rdob2}), implies the following decaying behavior for the matrix elements,
    \begin{equation}\label{eq:1rdmdecay}
    \rho^{(f)}_{k n}\leq const\times q(\kappa)^{n}\,,
    \end{equation}
    for $|k-n|< 2(N-1)$ and $|k-n|$ even. According to the previous two points all other matrix elements vanish.
\end{enumerate}

\begin{figure}
\includegraphics[width=8cm]{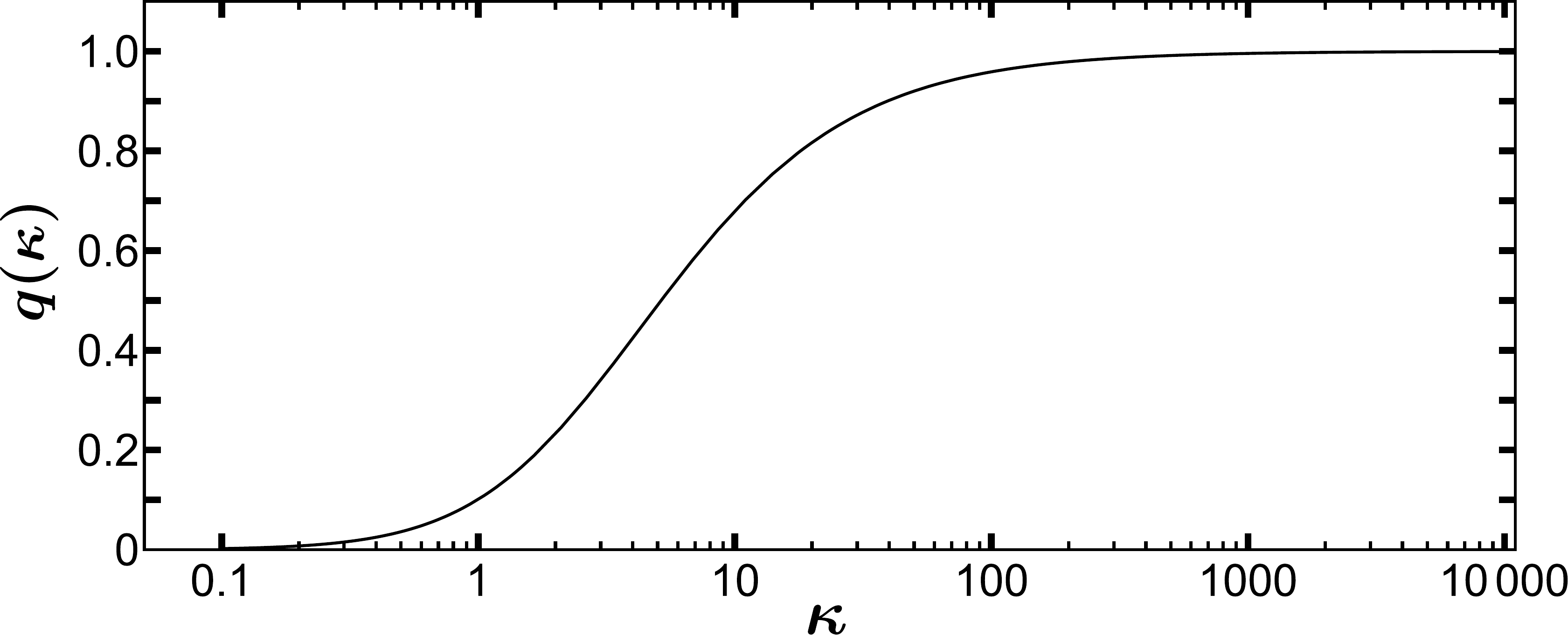}
\centering
\caption{Dependence of the decay constant $q(\kappa)$, defined in Eq.~\eqref{eq:qkappa}, on the interaction strength $\kappa$.}
\label{fig:q}
\end{figure}

The strong decaying hierarchy (\ref{eq:1rdmdecay}) of the matrix elements allows us to calculate numerically all relevant NONs for any fixed coupling strength $\kappa$ with very high precision. For this we truncate the corresponding infinite-dimensional 1-particle reduced density matrix to the left-upper  $R\times R$ block. For R sufficiently large the corresponding eigenvalues $\vec{\lambda}^{(R)}\equiv (\lambda_i^{(R)})_{i=1}^R$ are very close to the correct NONs. In particular, one can prove by using a norm estimate on the difference of spectra given the difference of the corresponding two matrices \cite{pertTheory} that
\begin{equation}\label{eq:NONdif}
\|\vec{\lambda}-\vec{\lambda}^{(R)}\|_1 \leq const\times q(\kappa)^{R+1}\,,
\end{equation}
where $\vec{\lambda}^{(R)}$ was extended to infinite dimensions by adding 0's.
Due to the significance of the decay constant $q(\kappa)$ we present it in Figure \ref{fig:q}
for a large coupling regime. The behavior of $q(\kappa)$ and the estimate (\ref{eq:NONdif}) guarantee that even the regime of ultra-strong
couplings can be treated numerically. For instance, in Ref.~\cite{parity} NONs were calculated numerically for couplings up to $\kappa=10^{12}$ .

For the regime of weak couplings we find
\begin{equation}\label{eq:qweak}
q(\kappa)= \frac{N-1}{N^2}\, \kappa^2+ \frac{N-1}{N^2}\, \kappa^3+ \mathcal{O}(\kappa^4)\,,
\end{equation}
which makes a perturbational approach for the weak coupling regime feasible.
To explain this, we expand the 1-particle reduced density matrix in a series,
\begin{equation}\label{eq:1rdmseries}
  \rho^{(f)}\equiv \sum_{n=0}^{\infty} \,\frac{1}{n!} \rho_n \,\kappa^n\,.
\end{equation}
The hierarchy (\ref{eq:1rdmdecay}) then implies that the corresponding degenerate Rayleigh-Schr\"odinger perturbation theory up to a fixed order $\kappa^s$ involves only matrices of finite (even very small) rank.

In Ref.~\cite{CS2013} such a perturbational approach was used for the case of $N=3$ and in the next section it will be used for $N=4$. It is worth noticing that such an approach can be significantly simplified by referring to a duality of NONs first observed in Ref.~\cite{Nagydual1} and proven in Ref.~\cite{duality}: $\vec{\lambda}(l/\tilde{l})= \vec{\lambda}(\tilde{l}/l)$ (recall Eq.~\eqref{eq:kappa}). By employing the alternative coupling parameter
\begin{equation}\label{eq:delta}
\delta:=\ln\left(\frac{l}{\tilde l}\right) = \frac{1}{4}\ln\left(1+ \kappa\right)=\frac{1}{4}\kappa +\mathcal{O}(\kappa^2)\,,
\end{equation}
this duality reads $\vec{\lambda}(\delta)=\vec{\lambda}(-\delta)$. As a consequence,
the series expansions of various NONs simplifies since it contains even orders in $\delta$, only.

\section{Pinning analysis for Harmonium in 1d}\label{sec:1d}

\begin{table}
\arraycolsep=2.5pt\def\arraystretch{1.3}
$
\begin{array}{r|c|c|c|c||c|c|c|c|l}
\hline
\ldots&\lambda_{N-3}&\lambda_{N-2}&\lambda_{N-1}&\lambda_{N}&\lambda_{N+1}&\lambda_{N+2}&\lambda_{N+3}&\lambda_{N+4} & \ldots \\  \hline
\ldots &\kappa^8&\kappa^6&\kappa^4&\kappa^4&\kappa^4&\kappa^4&\kappa^6&\kappa^8 & \ldots \\ \hline
\end{array}
$
\caption{Leading order corrections to the values 1 and 0, respectively, for the decreasingly-ordered NONs $\lambda_{i}(\kappa)$ for the N-Harmonium ground state for small coupling $\kappa$.}
\label{tab:activespace}
\end{table}
We use the techniques discussed in Section \ref{sec:1rdm} to determine analytically for weak couplings and numerically for intermediate and strong couplings the NONs of the N-Harmonium ground state (\ref{eq:gsf}). Then, by using the concept of truncation as introduced in Section \ref{sec:truncation} we systematically explore the occurrence of quasipinning.

\subsection{Weak couplings}\label{sec:weak}
A first analysis of 3-Harmonium in one spatial dimension for the regime of weak interactions was presented in Ref.~\cite{CS2013}. It revealed the remarkable $\kappa^8$-quasipinning, i.e.~$D_{min}\sim \kappa^8$. It was also explained that this quasipinning is non-trivial since $D_{min}$ is by four orders in $\kappa$ smaller than the distance of $\vec{\lambda}(\kappa)$ to the Hartree-Fock point $\vec{\lambda}_{HF}$ (recall Figure \ref{fig:polytope}), behaving as $\mbox{dist}_1(\vec{\lambda}(\kappa),\vec{\lambda}_{HF})\sim \kappa^4$.
We explore whether such quasipinning occurs for larger particle numbers as well and quantify its non-triviality in a more elaborated way by using the Q-parameter.

As a first step beyond the work in Ref.~\cite{CS2013}, we consider the 4-Harmonium ground state.
By applying degenerate Rayleigh-Schr\"odinger perturbation theory we determine expansion series for the NONs. According to the remarks at the end of Section \ref{sec:1rdm} the choice of $\delta$ (recall Eq.~(\ref{eq:delta})) as coupling parameter
simplifies the perturbation theory. We obtain the following series expansions up to corrections of the \mbox{order $\mathcal{O}(\delta^{10})$:}
\begin{eqnarray}\label{eq:NON4}
\lambda_{1}(\delta) &=& 1-\frac{555}{65536} \delta^8 + \mathcal{O}(\delta^{10}) \nonumber \\
\lambda_{2}(\delta) &=& 1 - \frac{5}{64}\delta^6 + \frac{11735}{196608}\delta^8 + \mathcal{O}(\delta^{10}) \nonumber\\
\lambda_{3}(\delta) &=& 1-\frac{15}{64}\delta^4 + \frac{95}{256}\delta^6 - \frac{30387}{65536}\delta^8 + \mathcal{O}(\delta^{10}) \nonumber\\
\lambda_{4}(\delta) &=& 1-\frac{15}{64}\delta^4+\frac{75}{256}\delta^6-\frac{63281}{196608}\delta^8 + \mathcal{O}(\delta^{10}) \nonumber\\
\lambda_{5}(\delta) &=& \frac{15}{64}\delta^4 -\frac{75}{256}\delta^6 \frac{58361}{196608}\delta^8+ \mathcal{O}(\delta^{10})\nonumber\\
\lambda_{6}(\delta) &=& \frac{15}{64}\delta^4 -\frac{95}{256}\delta^6 + \frac{30987}{65536}\delta^8+ \mathcal{O}(\delta^{10})\nonumber\\
\lambda_{7}(\delta) &=& \frac{5}{64}\delta^6 - \frac{15455}{196608}\delta^8+ \mathcal{O}(\delta^{10})\nonumber\\
\lambda_{8}(\delta) &=& \frac{2835}{65536}\delta^8+ \mathcal{O}(\delta^{10}) \nonumber\\
\lambda_{9}(\delta) &=& \mathcal{O}(\delta^{10})\,.
\end{eqnarray}
We have worked out such expansion series for all ground states up to $N=8$ particles.

For $3\leq N\leq8$ a well-pronounced hierarchy of active spaces occurs: By considering the scale $\mathcal{O}(\delta^4)$ always two NONs ($\lambda_{N-1},\lambda_N$) have corrections to the value 1 and two NONs ($\lambda_{N+1},\lambda_{N+2}$) to 0, respectively. On the finer scale $\mathcal{O}(\delta^6)$, also the NONs $\lambda_{N-2}$ and $\lambda_{N+3}$ deviate from their zero-interaction values. This hierarchy continues in that systematic way to higher orders in $\delta$: Whenever two more orders in $\delta$ are taken into account two additional NONs begin to deviate from 1 and 0, respectively. According to the linear leading order relation between $\delta$ and $\kappa$ (\ref{eq:delta}) the same hierarchy is found by referring to the more intuitive coupling strength $\kappa$. This hierarchy is also illustrated in Table \ref{tab:activespace}.

In the following, we exploit the concept of truncation, as developed in Section \ref{sec:truncation} for a systematic (quasi)pinning analysis of the NONs (\ref{eq:NON4}). The given hierarchy will simplify this task significantly.
\begin{itemize}
\item On the scale $\mathcal{O}(\delta^4)$, all NONs except $\lambda_3,\ldots,\lambda_6$ are identical to 1 and 0, respectively, and can therefore be neglected. Yet since the remaining setting $(2,4)$ is still trivial in the sense that the GPCs do not take the form of proper inequalities, the corresponding pinning analysis is meaningless \cite{CSQuasipinning}.
\item By considering the next significant scale, $\mathcal{O}(\delta^6)$, we obtain the truncated setting $(3,6)$. This so-called Borland-Dennis setting \cite{Borl1972} has the following GPCs (to avoid any confusion with the NONs in \eqref{eq:NON4} we denote the NONs by $\lambda'_i$, $i=1,2,\ldots,6$):
\begin{eqnarray}
  && \lambda'_1 +\lambda'_6= \lambda'_2 +\lambda'_5=\lambda'_3 +\lambda'_4=1\,, \label{eq:gpc36a} \\
  && D^{(3,6)}(\vec{\lambda}')\equiv2-(\lambda'_1 +\lambda'_2+\lambda'_4)\geq 0\,. \label{eq:gpc36b}
\end{eqnarray}
As a consistency check we observe $\lambda_i(\delta)+\lambda_{9-i}(\delta)=1+\mathcal{O}(\delta^8)$ for $i=2,3,4$. This means that the GPCs (\ref{eq:gpc36a}) are indeed fulfilled up to the truncation error, which is of the order $\mathcal{O}(\delta^8)$.
Only for GPC (\ref{eq:gpc36b}) the question of (quasi)pinning is meaningful. We find
\begin{equation}\label{eq:D36NON4}
D^{(3,6)}\big((\lambda_i(\delta))_{i=2}^7\big) =  -\frac{47569}{65536} \delta^8\,.
\end{equation}
Hence, GPC (\ref{eq:gpc36b}) is pinned up to corrections of the same order than the truncation error.
According to the concept of truncation this implies that the full spectrum of NONs is saturating at least some GPCs within the correct setting $(4,\infty)$ up to corrections of order $r=8$ or larger, $D_{min}\sim\delta^r$. To explore whether the quasipinning of the NONs is stronger or turns even into pinning we need to consider a finer scale.
\item On the  scale $\mathcal{O}(\delta^8)$ the truncated setting increases to $(4,8)$. There are $14$ GPCs listed in Appendix \ref{app:gpc48}, all taking the form of proper inequalities. The corresponding truncation error is given by $\mathcal{O}(\lambda_9)=\mathcal{O}(\delta^{10})$. The pinning analysis yields that all $14$ GPCs are saturated up to corrections of order $\delta^{8}$. Since this is larger by a factor $1/\delta^2$ than the truncation error  this already completes the (quasi)pinning analysis for the regime of not too strong couplings: The NONs of the 4-Harmonium ground states are not pinned. Yet they show surprisingly strong quasipinning of the order $D_{min}\sim \delta^8$.
\end{itemize}

Similar (quasi)pinning analyses for various particle numbers up to $N=8$ show that the NONs of the corresponding N-Harmonium ground state are strongly quasipinned in the regime of weak couplings.
In detail, we find for $4\leq N\leq 8$ (recall Eq.~(\ref{eq:delta}))
\begin{eqnarray}\label{eq:DNONN}
D_{min} &=& c_N \,\delta^{2N}+\mathcal{O}(\delta^{2N+2}) \nonumber \\
& =& d_N\, \kappa^{2N}+\mathcal{O}(\kappa^{2N+1})\,,
\end{eqnarray}
with $d_N=\frac{c_N}{4^{2N}}$. Recall that for $N=3$ one has $D_{min}\sim \delta^8$ \cite{CS2013}. There is little doubt that result (\ref{eq:DNONN}) holds for $N>8$ as well.

It is particularly remarkable that by adding another particle to the system the quasipinning becomes stronger by two additional orders in $\delta$ and $\kappa$, respectively. This, as well as the hierarchy of active spaces shown in Table \ref{tab:activespace}, expresses the existence of a kind of a `microscopic Pauli pressure'. This pressure built up by the additional particles is pressing $\vec{\lambda}$ closer to the boundary of the polytope $\mathcal{P}$ and the Pauli simplex $\Sigma$, respectively.

Besides result (\ref{eq:DNONN}) on the \emph{absolute} significance of the GPCs for the Harmonium ground state we also need to explore its relative significance, as it was explained in Section \ref{sec:Q}: To which extend can quasipinning (\ref{eq:DNONN}) by GPCs be deduced from possible quasipinning by Pauli exclusion principle constraints? Since the distance of $\vec{\lambda}(\delta)$ (\ref{eq:NON4}) to the Hartree-Fock-point $\vec{\lambda}_{HF}\equiv(1,\ldots,1,0,\ldots)$ is of the order $\delta^4$ at least four orders of the quasipinning $D_{min}\sim \delta^8$ are trivial. They already follow from weak correlations and thus from quasipinning by Pauli exclusion principle constraints.
A thorough analysis of the extend of trivial quasipinning in form of the elaborated Q-parameter (recall Section \ref{sec:Q}), however, implies
\begin{equation}\label{eq:Qweak}
Q\sim 2\log_{10}{\delta}\,.
\end{equation}
The same result is found for various N that were considered in this work, namely $3\leq N\leq 8$. This means that the strong quasipinning (\ref{eq:DNONN}) is non-trivial by two orders in $\delta$ and $\kappa$, respectively.

\subsection{Intermediate and strong couplings}\label{sec:interm}
The results on quasipinning for the 3-Harmonium ground state in Ref.~\cite{CS2013} and for the next few particle numbers $N=4,\ldots,8$, derived and discussed in the previous section, concern the regime of weak interaction. The series expansions, e.g., Eq.~(\ref{eq:NON4}), are valid as long as the mathematically more convenient coupling parameter (\ref{eq:delta}) is sufficiently small, $\delta\ll 1$ \footnote{In principle, the radius of convergence may be larger, and even extend to $\mathcal{O}(1)$. Yet we could not gain any profound insights on that.}. To determine the corresponding regime for the physically more significant coupling $\kappa$ (\ref{eq:kappa}) we observe that $\kappa(\delta=1)\approx 53.6$. Thus $\delta \ll 1$ covers indeed the whole regime of weak physical couplings, i.e.~$\kappa \lesssim 1$. In this section we extend the (quasi)pinning analysis also to intermediate and even strong and ultra strong coupling strengths. For this, we use the numerical tools discussed in Section \ref{sec:1rdm}. Since increasing the coupling between the fermions also leads to an increase of the correlations and particularly of the required dimension of the truncated 1-particle Hilbert space we need to carefully keep track of the truncation error as well.

As a first example, we explore pinning for the 1-dimensional harmonic analogue of the lithium atom, i.e.~Hamiltonian (\ref{eq:HamiltonianHarmonium01}) for $N=3$ with $3K=m \omega^2$ (implying $\kappa=1$). The most relevant NONs are listed in Table \ref{tab:LithiumAnalogHarm}.
\begin{table}
\begin{tabular}{|c|c||c|c|}
\hline
$\lambda_1$ & $0.999998533821$ & $\lambda_6$ & $0.000001455434$ \\ \hline
$\lambda_2$ & $0.999807955780$ & $\lambda_7$ & $0.000000028353$ \\ \hline
$\lambda_3$ & $0.999806535259$ & $\lambda_8$ & $0.000000000265$ \\ \hline
$\lambda_4$ & $0.000193443778$ & $\lambda_9$ & $0.000000000002$ \\ \hline
$\lambda_5$ & $0.000192047307$ & $\ldots$ & \ldots \\ \hline
\end{tabular}
\caption{There are shown all NONs $\lambda_i\geq 10^{-12}$ of the \mbox{3-Harmonium} ground state in one spatial dimension with particle interaction strength $\kappa=1$.}
\label{tab:LithiumAnalogHarm}
\end{table}
Very similar to the electrons in an atom the trapping potential makes this a weakly correlated system: All NONs are close to either 1 or 0. Due to the decaying hierarchy of NON which is still well pronounced for $\kappa=1$ we can still follow the same procedure as in Section \ref{sec:weak}, namely to consider successively different scales. The final result on quasipinning for the exact ground state is found as
\begin{equation}\label{eq:DminLi}
D_{min} = 7.66\times 10^{-9}\,.
\end{equation}
Note, that $D_{min}$ is of the same order as $\lambda_7$ and $\lambda_8\ll D_{min}$. This implies that such quasipinning (\ref{eq:DminLi}) can conclusively be confirmed already within the truncated setting $(3,7)$. To elaborate on its non-triviality we observe that the distance of $\vec{\lambda}$ to the Hartree-Fock point is given by
\begin{equation}\label{eq:DHfLi}
\mbox{dist}_1(\vec{\lambda},\vec{\lambda}_{HF}) \equiv \sum_{i=1}^3 (1-\lambda_i)+\sum_{j=4}^\infty \lambda_j= 7.74\times 10^{-4}\,.
\end{equation}
The distance $D_{min}$ of $\vec{\lambda}$ to the polytope boundary is smaller by about a factor $10^5$. Yet to confirm the non-triviality of quasipinning (\ref{eq:DminLi}) we resort to the overall Q-parameter \cite{CSQ} and eventually find $Q=1.85$. This means that $\vec{\lambda}$ is closer by a factor $10^{1.85}\approx71$ to some facets of the polytope $\mathcal{P}$ than one can expect from small distances of $\vec{\lambda}$ to the facets of the Pauli simplex (\ref{eq:Sigma}). In other words, the quasipinning by GPC is stronger by a factor $71$ than the quasipinning by PEP constraints.

\begin{figure}
\includegraphics[width = 8cm] {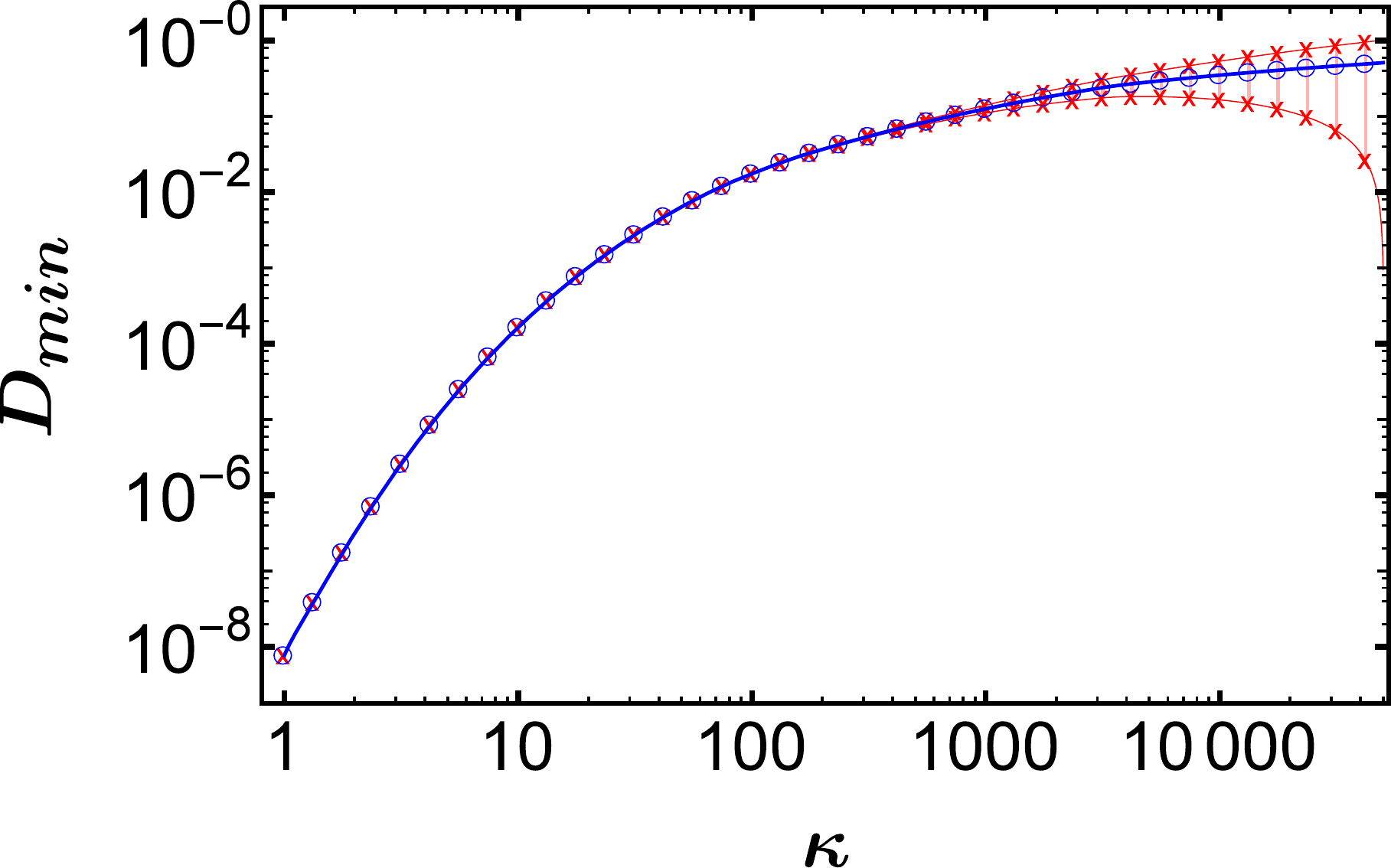}
\caption{Quasipinning exhibited by the 3-Harmonium ground state in one spatial dimension: Numerical data points of minimal $l^1$-distance $D_{min}$ of the NONs $\vec{\lambda}$ to polytope boundary $\partial \mathcal{P}$ for intermediate up to very strong interaction strengths (blue circles). Upper and lower error margins (red crosses) within the largest known setting $(N,d)=(3,11)$ are found negligible for even strong interactions up to $\kappa \approx 5000$.}\label{fig:3HarmoAndTruncError}
\end{figure}

In the same way as for $\kappa=1$ we now explore the occurrence of quasipinning and its non-triviality measured by the Q-parameter for various $\kappa$ quasi-continuously chosen within the regimes of intermediate, strong and ultra-strong couplings.
First, for the case of $N=3$ the results of the (quasi)pinning analysis are presented in Figure~\ref{fig:3HarmoAndTruncError}. The $\lambda$-vectors have been determined numerically for a set of logarithmically distributed $\kappa$-values. The associated quasipinnings $D_{min}$ have been interpolated and plotted in Figure \ref{fig:3HarmoAndTruncError} (blue circles), together with the related upper and lower error margins arising due to the truncation (red crosses) \footnote{We made a quite conservative and probably too pessimistic choice for the truncation error: We defined it by
\begin{equation}\Delta D_{min}=\sum_{i=1}^r(1-\lambda_i)+\sum_{j=d-s+1}^d\lambda_j\,,
\end{equation}
for the case that r NONs close to 1 and s NONs close to 0 were neglected. The definition of a truncation error is a bit arbitrary as long as it is not known yet how the coefficients in a GPC (\ref{eq:gpc}) could grow by adding more dimensions to the 1-particle Hilbert space.}. In particular, this also demonstrates that the concept of truncation (given the GPCs for $N=3$ up to dimension $d=11$) allows us to conclusively explore and quantify quasipinning even for very strong couplings, up to $\kappa\lesssim 5000$. For $\kappa \gtrsim 50000$ the truncation error becomes larger than the value for $D_{min}$ found in the truncated setting $(3,11)$ and a pinning analysis becomes meaningless.

The results for the 3-Harmonium ground state show that the existence of quasipinning extends to intermediate and even strong couplings. Yet it also reduces the more we increase $\kappa$. Only in the regime of quite strong couplings the quasipinning vanishes and the distance of $\vec{\lambda}(\kappa)$ to the polytope boundary reaches the scale of the diameter of the polytope, i.e.~$\mathcal{O}(1)$.

For the next larger particle numbers $4\leq N\leq 8$ we find qualitatively similar quasipinning curves $D_{min}(\kappa)$ as for $N=3$. For the regime of weak and up to intermediate couplings of $\mathcal{O}(1)$, the behavior of $D_{min}(\kappa)$ is described by the perturbational result (\ref{eq:DNONN}). Yet for $\kappa \gtrsim 10$ all curves $D_{min}(\kappa)$ begin to approach each other more and more, the quasipinning reduces and eventually vanishes for very strong couplings.

\begin{figure}[]
\centering
\includegraphics[width=8.0cm]{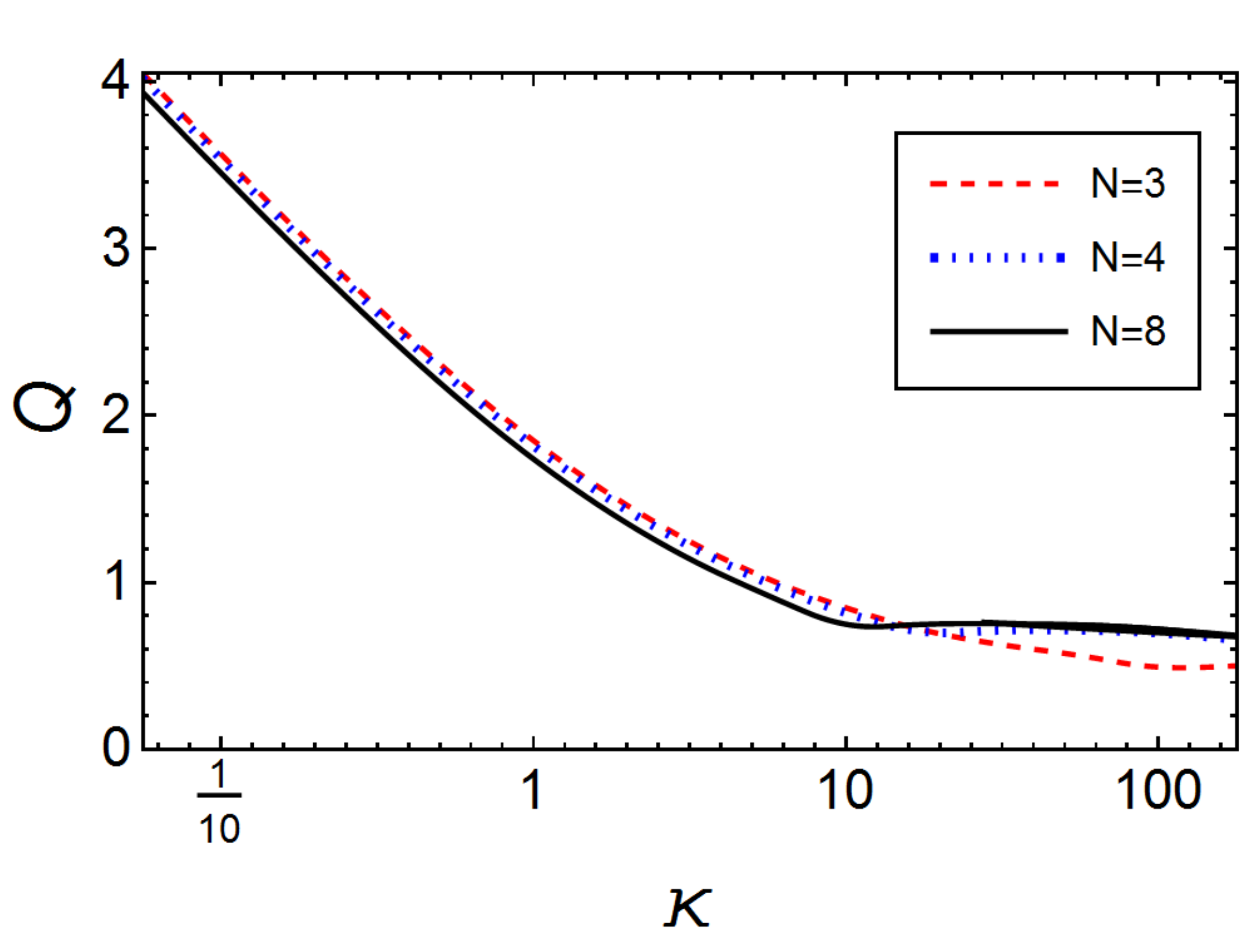}
\caption{Non-triviality of quasipinning: Q-measure for the N-Harmonium ground state with $N=3,5,8$ as function of the coupling $\kappa$. $Q$ is seen to be similar for various particle numbers. The logarithmic plot reveals a linear behavior for small interaction strengths $\kappa$. As the corresponding slope is 2, this confirms that the quasipinning is non-trivial by two orders in $\kappa$.}
\label{fig:N1dQ}
\end{figure}

The non-triviality of the quasipinning is explored and quantified again by the Q-parameters. The corresponding results are shown in Figure \ref{fig:N1dQ}. $Q(\kappa)$ is very similar for all $N=3,\ldots,8$. It decreases with increasing coupling $\kappa$ and one finds $Q\sim -2 \log_{10}(\kappa)$, independent of N in the regime of weak couplings in agreement with the analytic result, Eq.~(\ref{eq:Qweak}). This proves that in comparison to the approximate saturation of Pauli exclusion principle constraints, the distance between the $\lambda$-vector and the polytope boundary is smaller by two orders in $\kappa$. The results in Figure \ref{fig:N1dQ} further demonstrate that quasipinning remains non-trivial up to medium interaction strengths but becomes quite trivial for larger couplings.

\section{Summary}\label{sec:concl}
The fermionic exchange symmetry implies restrictions on occupation numbers stronger than Pauli's exclusion principle. Those generalized Pauli constraints (GPCs) restrict the vector $\vec{\lambda}\equiv(\lambda_i)_{i=1}^d$ of natural occupation numbers (NONs), the eigenvalues of the 1-particle reduced density operator, to a polytope $\mathcal{P}\subset [0,1]^d$. Physical significance of GPCs is particularly given whenever the vector $\vec{\lambda}$ of NONs is found on (\emph{pinning}) or at least very close (\emph{quasipinning}) to the polytope boundary $\partial \mathcal{P}$.

We have provided a conclusive analysis of the occurrence of pinning and quasipinning for a 1-dimensional few-fermion quantum system.

We have first elaborated on measures allowing us to quantify quasipinning. Although measures given by the $l^p$-distance  $\mbox{dist}_p(\vec{\lambda},\partial \mathcal{P})$ have some significance for all $p$, it is explained that the 1-norm is the most preferable one. We have introduced and explained the concept of truncation which allows one to simplify the analysis of possible quasipinning by skipping various NONs sufficiently close to 1 and 0. This is of practical importance since most few-fermion models are typically based on very large (even infinite) dimensional 1-particle Hilbert spaces but the families of GPCs are known so far only up to dimension $d=11$.

To explore the occurrence of pinning and quasipinning we have thoroughly studied the few-fermion  model system Harmonium (\ref{eq:HamiltonianHarmonium01}). In detail, we have explained how to determine the natural occupation numbers for the ground states for various particle numbers: For the regime of weak couplings we have resorted to degenerate Rayleigh-Schr\"odinger perturbation theory and for intermediate up to ultra-strong couplings we have used an exact numerical diagonalization.

By applying the concept of truncation to the infinite vector $\vec{\lambda}(\kappa)$ of NONs we succeeded in conclusively exploring quasipinning and in providing further evidence against the existence of the pinning-effect. For the regime of weak interactions, we have confirmed the existence of quasipinning reported for $N=3$ in Ref.~\cite{CS2013} also for the next larger particle numbers up to $N=8$. It turns out that this quasipinning behaves as $D_{min} \sim \kappa^{2N}$ for $4\leq N\leq 8$ which likely may extends to $N>8$ as well. This especially means that quasipinning becomes even stronger for larger particle numbers.
We speculate that this may be understood as a consequence of a \emph{`microscopic Pauli pressure'} built up from the particles and `pressing' $\vec{\lambda}$ closer to the polytope boundary.

Independent of N, we have also found that quasipinning extends to the regime of intermediate and also strong couplings but vanishes for ultra strong interaction strengths.

Since the GPCs are more restrictive than the Pauli exclusion principle (recall Figure \ref{fig:polytope}) quasipinning by GPCs can be a consequence of quasipinning of Pauli exclusion principle constraints (as e.g.~in case of weak correlations). By employing the recently developed Q-parameter \cite{CSQ} we have systematically explored and quantified such potential trivialities of quasipinning. The investigation proves that quasipinning by GPCs in the regime of weak couplings $\kappa$ and for all considered $N\leq 8$ is non-trivial by two orders in $\kappa$. This means that the distance of $\vec{\lambda}$ to the polytope boundary is by two orders in $\kappa$ smaller than one would expect from the approximate saturation of some Pauli exclusion principle constraints. In the regime of strong couplings the quasipinning becomes trivial.

We thank M.\hspace{0.5mm}Christandl, D.\hspace{0.5mm}Gross, D.\hspace{0.5mm}Jaksch and B.\hspace{0.5mm}Yadin for helpful discussions. We gratefully acknowledge financial support from
the Friedrich-Naumann-Stiftung and Christ Church Oxford (FT), the Foundational Questions Institute (FQXi-RFP3-1325) and the HKU Seed Funding for Basic Research (DE), the Oxford Martin School, the NRF (Singapore), the MoE (Singapore) and the EU Collaborative Project TherMiQ (Grant Agreement 618074) (VV), the Swiss National Science Foundation (Grant P2EZP2 152190) and the Oxford Martin Programme on Bio-Inspired Quantum Technologies (CS).

\bibliography{bibliography,comments}

\onecolumngrid
\appendix
\section{Generalized Pauli constraints for the setting $(4,8)$}\label{app:gpc48}
For the setting $(N,d)=(4,8)$ there are $14$ generalized Pauli constraints (see Ref.~\cite{Altun}):
\begin{eqnarray}\label{eq:gpc48}
 D_{1}^{(4,8)}(\vec{\lambda}):=& \quad 0 - \lambda_5 + \lambda_6 + \lambda_7 + \lambda_8 \geq 0  \\
D_{2}^{(4,8)} (\vec{\lambda}):=& \quad 0 - \lambda_1 + \lambda_2 + \lambda_7 + \lambda_8 \geq 0  \\
D_{3}^{(4,8)}(\vec{\lambda}) :=& \quad 0 - \lambda_1 + \lambda_3 + \lambda_6 + \lambda_8 \geq 0  \\
D_{4}^{(4,8)}(\vec{\lambda}) :=& \quad 0 - \lambda_1 + \lambda_4 + \lambda_6 + \lambda_7 \geq 0  \\
D_{5}^{(4,8)}(\vec{\lambda}) :=& \quad 0 - \lambda_1 + \lambda_4 + \lambda_5 + \lambda_8 \geq 0  \\
D_{6}^{(4,8)}(\vec{\lambda}) :=& \quad 0 - \lambda_3 + \lambda_4 + \lambda_7 + \lambda_8 \geq 0  \\
D_{7}^{(4,8)}(\vec{\lambda}) :=& \quad 0 - \lambda_2 + \lambda_4 + \lambda_6 + \lambda_8 \geq 0  \\
D_{8}^{(4,8)}(\vec{\lambda}) :=& \quad 2 - \lambda_2 - \lambda_3 - \lambda_5 + \lambda_8 \geq 0  \\
D_{8}^{(4,8)}(\vec{\lambda}) :=& \quad 2 - \lambda_1 - \lambda_3 - \lambda_6 + \lambda_8 \geq 0  \\
D_{10}^{(4,8)}(\vec{\lambda}) :=& \quad 2 - \lambda_1 - \lambda_2 - \lambda_7 + \lambda_8 \geq 0  \\
D_{11}^{(4,8)}(\vec{\lambda}) :=& \quad 2 - \lambda_1 - \lambda_2 - \lambda_3 + \lambda_4 \geq 0  \\
D_{12}^{(4,8)}(\vec{\lambda}) :=& \quad 2 - \lambda_1 - \lambda_4 - \lambda_5 + \lambda_8 \geq 0  \\
D_{13}^{(4,8)}(\vec{\lambda}) :=& \quad 2 - \lambda_1 - \lambda_2 - \lambda_5 + \lambda_6 \geq 0  \\
D_{14}^{(4,8)}(\vec{\lambda}) :=& \quad 2 - \lambda_1 - \lambda_3 - \lambda_5 + \lambda_7 \geq 0
\end{eqnarray}

\end{document}